\documentclass[reqno]{amsart}
\usepackage{amsfonts,amsmath} 
\date{Aug. 4, 2000} 
\input epsf     

\def\ip#1{\left < #1 \right >}

\def\dim{\operatorname{dim}}  

\def\Z{{\mathbb Z}}    
\def\R{{\mathbb R}}    
\def\C{{\mathbb C}}    

\def\O{{\rm O}}

\def\1{{\bf 1}}
\newtheorem{thm}{Theorem}[section]

\newtheorem{prop}[thm]{Proposition}
\theoremstyle{definition}

\theoremstyle{remark}

\numberwithin{equation}{section}

\newcommand{\eps}{\varepsilon}

\def\blackbox{{\vrule height 1.3ex width 1.0ex depth -.2ex}
       \hskip 1.5truecm}
\newenvironment{proof_of}[1]{
    \noindent{\bf Proof of #1: } }{
    \hfill \blackbox \bigskip}

\begin{document}
\bibliographystyle{mabib}
\title[Spectral Gaps and Graph Decoration]{The Creation of Spectral 
Gaps by Graph Decoration} 
\author[J.H. Schenker]{Jeffrey H. Schenker$^{(a)}$}
\address{Princeton University, Departments of Mathematics$^{(a)}$ 
and Physics$^{(b)}$, Princeton, NJ 08544, USA.}
\email{schenker@princeton.edu, aizenman@princeton.edu}
\author[M. Aizenman]{Michael Aizenman$^{(a,b)}$} 
\begin{abstract}
We present a mechanism for the creation of gaps in the 
spectra of self-adjoint operators defined over a Hilbert space of 
functions on a graph, which is based on  the process of 
graph decoration.  The resulting Hamiltonians 
can be viewed as associated with discrete models exhibiting a repeated 
local structure and a certain  bottleneck in the hopping amplitudes.  
\end{abstract}
\maketitle
\section{Introduction}
Energy spectra characterized by the presence of bands and gaps are
familiar from the Bloch theory of periodic systems. In this note,
we present another mechanism for the creation of spectral gaps 
which does not rely on translation invariance.

The band-gap spectral structure plays an important role in the 
theory of the solid state~\cite{Kittel}, as well as in the properties 
of dialectric and acustic media~\cite{FiKl}.  
Of particular interest are also situations in which localized 
states are injected into existing gaps (see ref.~\cite{DH} 
 for a mathematical discussion with further references).  
These applied models are mentioned only as distant analogies; 
the topic we discuss pertains to spectral properties of Hamiltonians 
of discrete models, whose ``hopping terms'' can be viewed as 
associated with graphs exhibiting a repeated local structure 
and a certain  bottleneck in hopping amplitudes.   

\begin{figure}[tbp]
\leavevmode
\centering \epsfxsize=3in \epsffile{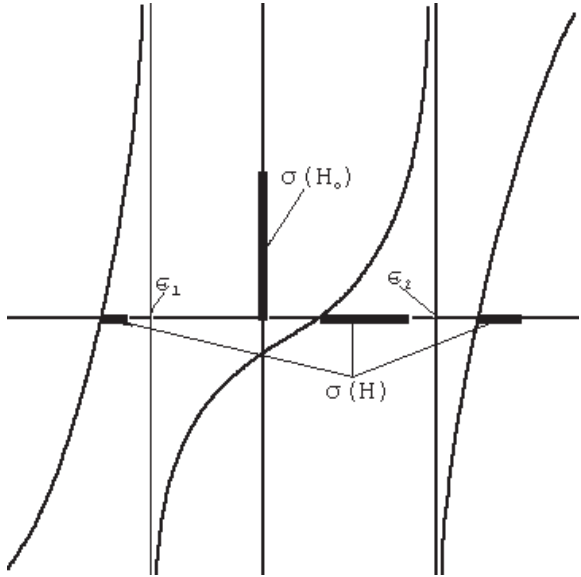}
\caption{\label{fig:typicalgamma} A schematic depiction of 
the spectral relation  $\sigma(H) = \gamma^{-1}(\sigma(H_{o}) )$. }
\end{figure}

To present the principle described herein it is convenient to 
introduce the notion of ``graph decoration''.
Given two graphs $\Gamma$ and $G$, we may ``decorate''
$\Gamma$ with $G$ by ``gluing'' a copy of $G$ to each vertex $v$
of $\Gamma$ in such a way that $v$ is identified with the
appropriate copy of some distinguished vertex $\O_G \in G$ (see
\S\ref{sec:decoration} for a formal definition, and figures
\ref{fig:example1} and \ref{fig:example2} for typical examples).
Given self-adjoint operators $H_o$ on $\ell^2(\Gamma)$ and $A$ on
$\ell^2(G)$ there is a natural way to define an operator extension
$H$ of $H_o$ and $A$ to $\ell^2(\Gamma \triangleleft G)$ where
$\Gamma \triangleleft G$ denotes the decorated graph just
described. In the absence of certain degeneracy, there is a simple
relation between the spectra of $H$ and $H_o$, denoted here by  
$\sigma(H_{...})$,  
which allows us to
conclude that intervals around certain energies $\eps_j$ are
excluded from the spectrum of $H$.
 Specifically, there is a
function $\gamma$ such that 
\begin{equation}
  \sigma (H) \ = \ \gamma^{-1} \left ( \sigma (H_o) \right ) \; ,
\label{eq:spectralsplitting}
\end{equation}
and $\gamma$ is of the form
\begin{equation}
  \gamma(E) \ = \ E + c + \sum_j w_j \ {1 \over \eps_j - E} \; ,
\end{equation}
where $w_j > 0$ and $\eps_j,c \in \R$ (see figure
\ref{fig:typicalgamma}).
In fact, we shall see that $\eps_j$ are 
exactly the eigenvalues of the operator 
$\widehat P A \widehat P $ where $\widehat P$ is the projection
onto the subspace of functions in $\ell^{2}(G)$ which vanish at
$\O_{G}$.
Thus, we have the 
 appealing picture in which the eigenenergies of the
decorated graph are repelled by
resonances with the ``inner spectrum'' of the decoration.

In \S\ref{sec:decoration} we define graph decoration and describe
the operator extension mentioned above. In \S\ref{sec:evaluation}
we present our main result (Prop. \ref{prop:main}) which describes
the spectral relationship presented above in full detail. Finally,
in \S\ref{sec:examples} we provide several examples and
applications of Prop. \ref{prop:main}.

\section{Graph decoration}\label{sec:decoration}

In this section, we suppose that we are given a graph $\Gamma$ and
a self adjoint operator $H_o$ on $\ell^2(\Gamma)$, the space of
square summable functions on the vertices of $\Gamma$ (see below).
Our goal is to describe a certain class of graph extensions of
$\Gamma$ and a corresponding class of operator extensions of
$H_o$.

Recall that a graph $G$ is described by specifying two sets: (1)
$V(G)$ whose elements are called vertices, and (2) $E(G)$ a set of
(unordered) pairs of vertices called edges. The edges play a
secondary role in our discussion, for we are mainly concerned with
the Hilbert space of square summable functions mapping $V(G)
\rightarrow \C$, which we denote $\ell^2(\Gamma)$.  The situation
of interest is when $E(G)$ is defined so that a given operator $A$
on $\ell^2(G)$ is {\it compatible} with $G$, by which we mean that
the off diagonal matrix elements $\ip{x|A|y}$ vanish whenever
$\{x,y\} \not \in E(G)$.
\footnote{We use the Dirac bra-ket notation for
matrix elements in the standard basis
$  \left | x \right > \ = \ \delta_x  $, with 
$\delta_x (y) $ the Kronecker function.}
Correspondingly, our notation generally identifies a graph $G$
with its vertex set: by $x \in G$ we indicate $x \in V(G)$ and we
shall say that a graph is countable (finite) if $V(G)$ is
countable (finite).

The graph extensions of $\Gamma$ shall be obtained by ``gluing''
copies of a second graph $G$ to each vertex of $\Gamma$. The
extended graph may be visualized as a field in which are tethered
many identical kites (see figures \ref{fig:example1} and
\ref{fig:example2}). Formally given any graph $G$ with a
distinguished vertex $\O_G$ we define the {\it decoration of~
$\Gamma$ by $(G,\O_G)$}, denoted $\Gamma \triangleleft G$, to be
the following graph:
\begin{enumerate}
\item $V \left ( \Gamma \triangleleft G \right ) = V(\Gamma) \times V(G)$ .
\item $E \left ( \Gamma \triangleleft G \right ) =
    E_{\rm field} \cup E_{\rm kite}$ , where:
\begin{enumerate}
\item $E_{\rm field}= \left \{ \ \{(x,\O_G),(y,\O_G)  \} \ | \
\{x,y\} \in E(\Gamma) \ \right \}$ .
\item $E_{\rm kite} = \left \{ \  \{(x,h),(x,g)\} \ | \ x \in V(\Gamma)
\mbox{ and
} \{h,g\} \in E(G) \ \right \}$ .
\end{enumerate}
\end{enumerate}

We think of the space  $\ell^2(\Gamma \triangleleft G)$  as the tensor
product  $\ell^2(\Gamma) \otimes \ell^2(G)$, which is natural
since the vertex set of $\Gamma \triangleleft G$ is $V(\Gamma) \times
V(G)$.
The subspace of functions which are supported on $\Gamma_o =
\{(x,\O_g): x \in \Gamma\}$ is naturally identified with
$\ell^{2}(\Gamma)$.  We denote by $P$  the orthogonal
projection onto this space.

Let $A$ be a self adjoint operator
on $\ell^2(G)$.  A natural extension of $H_{o}$ to
$\ell^2(\Gamma) \otimes \ell^2(G)$, incorporating  $A$, is
\begin{equation}\label{eq:defnofH}
  H \ := \ P H_o P + \1 \otimes A \;.
\end{equation}
The above operator is appropriate to the geometry of
graph decoration, for if $H_o$ and $A$ are compatible with $\Gamma$
and $G$ respectively, then $H$ is compatible with $\Gamma
\triangleleft G$.

An example of an operator of the form described in eq.
\eqref{eq:defnofH} is provided by the discrete Laplacian. On any
graph $H$ the discrete Laplacian, $\Delta_H$, is defined by
\begin{equation}
  [\Delta_H \psi](x) \ := \ \sum_{y:\, \{x,y\} \in E(H)} \psi(y) -
  \psi(x) \; .
\end{equation}
For decorated graphs, if we take $H_o = -\Delta_\Gamma$ and $A =
- \Delta_G$ then the operator defined by \eqref{eq:defnofH} is  
$H = -\Delta_{\Gamma \triangleleft G}$.

\section{A resolvent evaluation principle}\label{sec:evaluation}
We now focus on the case $|G| < \infty$ and present our main
result.\footnote{This result may be easily extended to the case
$|G|= \infty$ provided the spectrum of $A$ is discrete.}

\begin{prop}\label{prop:main}
Let $H$ be a bounded \footnote{More generally, $H$ may be
unbounded provided the set of functions with finite support forms a
core for $H$.} self adjoint operator of the form described in eq.
\eqref{eq:defnofH} with $G$ a finite graph. If $\left |\O_G \right
> $ is a cyclic vector for $A$, then
\begin{equation}\label{eq:spectralmap}
\sigma(H)  \ = \   \gamma^{-1}\left (\sigma(H_o) \right )\; ,
\end{equation}
where $\gamma$ is a function of the form
\begin{equation}\label{eq:formofgamma}
  \gamma(E)
     \ = \ E \ + \ c \ + \sum_{j=1}^{n} w_j  {1 \over \eps_j - E } \; ,
\end{equation}
with $c,\eps_j \in \R$, and $w_j > 0$.

Furthermore, whether or not $\left |\O_G \right>$ is cyclic, there
is a function $\gamma$ of the form \eqref{eq:formofgamma} such
that for each $x \in \Gamma$ and $z \in \C \setminus \R$
\begin{equation}\label{eq:greenrelation}
  \ip{x,\O_G| (H-z)^{-1} |x,\O_G} \ = \ \ip{x|(H_o-\gamma(z))^{-1}|x} \; ,
\end{equation}
and the spectral measure, $\widetilde \mu_x$,  for $H$ associated
to $\left |x,\O_G \right >$ is related to the spectral measure,
$\mu_x$, for $H_o$ associated to $\left |x \right >$ by
\begin{equation}\label{eq:spectralmeasures}
  \widetilde \mu_x(dE)  \ = \ {1 \over \gamma'(E)} \ \mu_x(d \gamma(E))  \;
  .
\end{equation}
Thus
\begin{equation}\label{eq:spectralsubset}
\gamma^{-1}(\sigma(H_o))  \ \subseteq \  \sigma(H) \; .
\end{equation}
\end{prop}

{\noindent \bf Remarks:}
\begin{itemize}
\item Recall that given a self-adjoint operator $K$ on a Hilbert space
$\mathcal K$, the spectral measure associated to a vector $v \in
{\mathcal K}$ is defined via the functional calculus and the
Riesz-Markov theoerem as the unique regular Borel measure, $\nu$,
such that
\begin{equation}
  \int f(E) \nu(dE) \ = \ \ip{v,f(K)v} \; ,
\end{equation}
for each $f \in C_o(\R)$, the family of continuous functions on
$\R$ which vanish at infinity.
\item Eq. \eqref{eq:spectralmeasures} is a formal expression which
indicates the following identity for the expectations of a
function $f \in C_o(\R)$:
\begin{equation}
    \int  \widetilde \mu_x(dE) \, f(E)  \ = \ \int \mu_x(d \eps) \,
    \sum_{E \in \gamma^{-1}(\eps)} {1 \over \gamma'(E)} f(E)  \; .
\end{equation}
\end{itemize}

\begin{proof_of}{Proposition \ref{prop:main}}
The heart of Prop.~\ref{prop:main} is the relation
\eqref{eq:greenrelation}, so let us begin with a derivation of
this equation.  Fix $x \in \Gamma$ and $z \in \C \setminus \R$.
Recall that the Green function,
\begin{equation}
  G(y,u) \ := \ \ip{x,\O_G | (H-z)^{-1} |y,u} \; , \  (y,u) \in
  \Gamma \triangleleft G \; ,
\end{equation}
is the unique square summable solution to the equation
\begin{equation}\label{eq:greeneq}
  (H - z )G = \left | x, \O_G \right > \; .
\end{equation}
A natural guess is that the solution  factors:
\begin{equation}\label{eq:greenansatz}
  G(y,u) \ = \ g(y) \, h(u) \; .
\end{equation}
With this ansatz, eq. \eqref{eq:greeneq} yields for $g$
and $h$:
\begin{subequations} 
\begin{equation}
 g(y) \, [ (A - z)h](u) \ = \ 0 \; , \ u \neq \O_G \; ,
\label{eq:g}
\end{equation}
and
\begin{equation}
  [H_o g](y) \, h(\O_g)  \ + \ g(y) \, [(A - z)h](\O_g)
  \ = \ \left | x \right > \; .
\label{eq:h}
\end{equation}
\end{subequations} 
It is now an easy exercise to solve these equations using the
Green functions for $H_o$ and $A$: eq. \eqref{eq:g}  
gives $g$
as a function of $h$,
\begin{equation}
  g(y) \ = \ {1 \over h(\O_G)} \, \ip{x| \left (H_o + { [(A -
  z)h](\O_g) \over h(\O_g) } \right )^{-1} |y} \; ;
\end{equation}
while eq. \eqref{eq:h}   
shows that $h$ is a multiple of the Green
function for $A$,
\begin{equation}
  h(u) \ = \ C \, \ip{\O_G | (A-z)^{-1}|u} \; ,
\end{equation}
with $C$ an arbitrary factor which drops out of the resulting
solution:
\begin{equation}
  G(y,u)\  \ = \ \ip{x| \left (H_o + { 1 \over \ip{\O_G | (A-z)^{-1}|\O_G} }
   \right )^{-1} |y} \, {\ip{\O_G | (A-z)^{-1}|u} \over \ip{\O_G |
   (A-z)^{-1}|\O_G}} \; .
\end{equation}
Setting $u = \O_G$ and $y = x$ in this expression yields
\eqref{eq:greenrelation} with
\begin{equation}
  \gamma(z) \ := \ {-1 \over \ip{\O_G | (A - z)^{-1} |\O_G} } \; .
\end{equation}

Because $\dim \ell^2(G)$ is finite, $\gamma$ is a rational
function with finitely many simple real poles (which occur at the
zeros of $\ip{\O_G | (A- E)^{-1} | \O_G}$).   Hence, the partial
fraction expansion
(alternatively, the  represenation theory of Herglotz
functions)
shows that $\gamma$ is of the form displayed in
eq. \eqref{eq:formofgamma}:
\begin{equation}
  \gamma(z) \ = \ z  \ - \ c \
     + \ \sum_{j=1}^{n -1} w_j {1 \over \eps_j - z} \; ,
\end{equation}
with $c, \eps_j \in \R$ and $w_j > 0$.  (The coefficient of $z$ is
one, since $z / \gamma(z) \rightarrow 1$ as $z \rightarrow
\infty$.)

We now turn to the verification of the relation between the
spectral measures expressed in \eqref{eq:spectralmeasures}.  First
consider the situation when $\Gamma$ is finite.  For a
self-adjoint operator $B$ on a finite dimensional vector space
there is a useful formula for the spectral measure, $\nu_\psi$,
associated to a vector $\psi$:
\begin{equation}
  \nu_\psi(dE) \ = \ \delta \left ( {-1 \over \ip{\psi,
  (B-E)^{-1} \psi}} \right )dE \; ,
\label{eq:deltaequation}
\end{equation}
where  $\delta$ is the Dirac-delta ``function.''
(This formula offers a simple route to  the
``spectral averaging'' principle discussed in ref.~\cite{SW};
its derivation
is an instructive exercise which we leave to the reader.)
Coupled with eq.~\eqref{eq:greenrelation},
\eqref{eq:deltaequation} easily
yields:
\begin{multline}
  \widetilde \mu_x(dE) \ = \ \delta \left
  ( {-1 \over \ip{x,\O_G| (H - E)^{-1}|x,\O_G}} \right
  ) dE \\ = \ \delta \left ( {-1 \over \ip{x|
  (H_o  - \gamma(E))^{-1}|x}} \right
  )  {1 \over \gamma'(E)} d\gamma(E) \ = \
  {1 \over \gamma'(E)} \mu_x(d\gamma(E)) \; .
\end{multline}

When $\Gamma$ is infinite we must turn to a more abstract
derivation of eq. \eqref{eq:spectralmeasures}. Writing each side
of eq. \eqref{eq:greenrelation} as a spectral integral we find
that
\begin{equation}
\int {1 \over E - z} d \widetilde \mu_x (E) \ = \  \int {1 \over E
- \gamma(z)} d\mu_x(E) \; .
\end{equation}
Expanding the right side of this equality with partial fractions
yields
\begin{equation}
  \int {1 \over E - z} d \widetilde \mu_x (E) \ = \  \int
  \sum_{\lambda \in \gamma^{-1}(E)} {1
  \over \gamma'(\lambda)} {1 \over \lambda - z} d\mu_x(E) \; .
\label{eq:rescase}
\end{equation}
This equation can be viewed as a special case of:
\begin{equation}
  \int f(E) \, d \widetilde \mu_x(E) \ = \ \int \sum_{\lambda \in
  \gamma^{-1}(E)}{1 \over \gamma'(\lambda)} f(\lambda) \, d \mu_x(E)
  \; ,
\label{eq:general}
\end{equation}
and indeed \eqref{eq:rescase} implies \eqref{eq:general},
for all $f \in \C_o(\R)$,  since
by the Stone-Weierstrass Theorem
the set of finite sums of the form $\sum_j c_j {1 \over
E - z_j}$ is dense in  $ \C_o(\R)$.  As mentioned previously,
\eqref{eq:general} is the statement claimed in
\eqref{eq:spectralmeasures}.

Finally, the spectral inclusion $\gamma^{-1}(\sigma(H_o))
\subseteq \sigma(H)$ follows since
\begin{equation}\label{eq:unionofsupports}
  \gamma^{-1}(\sigma(H_o)) \ = \ \bigcup_{x \in \Gamma}
  \mbox{supp.}(\widetilde \mu_x) \; ,
\end{equation}
which may be verified using \eqref{eq:spectralmeasures}.  If
$\{\left | x,\O_G \right >\}$ is a cyclic family for $H$ then eq.
\eqref{eq:unionofsupports} shows further that
$\gamma^{-1}(\sigma(H_o)) = \sigma(H)$.  In case $\left | \O_G
\right >$ is a cyclic vector for $A$, this family is easily seen
to be cyclic for $H$.  This completes the proof of the
proposition.
\end{proof_of}

We conclude this section with several remarks regarding
proposition \ref{prop:main} and its proof:

\begin{enumerate}
\item Eigenfunctions and generalized eigenfunctions of $H$ factor
in the same way as the Green function (see eq.
\eqref{eq:greenansatz}).  That is
\begin{equation}
  \Psi(y,u) \ = \ \psi(y) \, \phi(u)
\end{equation}
satisfies $H\Psi = E \Psi$ provided
\begin{subequations}
\begin{equation}
\phi(u) \ = \ {\ip{\O_G|(A-E)^{-1}|u} \over  \ip{\O_G|(A-E)^{-1}
|\O_G} } \; ,
\end{equation}
and
\begin{equation}
H_o \psi = \gamma(E) \psi \; .
\end{equation}
\end{subequations}

\item The relationship between the spectrum of $H_o$ and $H$ is
a stronger relationship than the simple inclusion
$\gamma^{-1}(\sigma(H_o)) \subset \sigma(H)$: {\em the spectral
type is preserved under the map $\gamma^{-1}$}.  So, a bound state
for $H_o$ gives rise to $n$ bound states for $H$. Similarly, a
band of absolutely continuous spectrum for $H_o$ gives rise to $n$
bands of absolutely continuous spectrum for $H$. If $H_o$
possesses singular continuous spectrum, then such spectrum also
occurs in the spectrum of $H$.

\item Generically, $\left |\O_G \right >$ is a cyclic vector for $A$, and
$\sigma(H) \ =  \ \gamma^{-1}(\sigma(H_o))$.  However, even when
$\left |\O_G \right >$ is not cyclic, we may still determine the
spectrum of $H$. We need only decompose the space $\ell^2(G)$ as a
direct sum $V \oplus V^\bot$ with each summand invariant under $A$
and $\left |\O_G \right >$ cyclic for $A|_V$. Then,
\begin{equation}
  \sigma(H) = \gamma^{-1}(\sigma(H_o)) \cup
  \sigma(A|_{V^\bot}) \; ,
\end{equation}
which may be verified by noting that
\begin{equation}
  H \cong \begin{pmatrix}
    H_o + \1 \otimes A|_V & 0 \\
    0 & \1 \otimes A|_{V^\bot}
  \end{pmatrix} \; .
\end{equation}
Note that the eigenvalues in $\sigma(A|_{V^\bot})$ occur with
multiplicity a multiple of $|\Gamma|$.
\item The poles $\eps_j$ of $\gamma$ are
eigenvalues of $\widehat A = \widehat P A \widehat P$, where
$\widehat P$ is the projection of $\ell^2(G)$ onto the space of
functions which vanish at $\O_G$.  To see this, recall that
$\eps_j$ satisfy $\ip{\O_G | (A-\eps_j)^{-1} |\O_G} = 0$. Thus the
Green functions $\psi_{j}(y) \ = \ \ip{\O_G | (A - \eps_j)^{-1}
|y}$ themselves are eigenfunctions: $\widehat A \psi_j = \eps_j
\psi_j$. Conversely, if $\left | \O_G \right > $ is cyclic for $A$ then
every eigenvalue of $\widehat A$ is a pole of $\gamma$.
\item  The coefficients $c$ and $w_j$
in the partial fraction expansion of $\gamma$ (eq.
\eqref{eq:formofgamma}) satisfy
\begin{subequations}
\begin{equation}
  c \ = \ -\ip{\O_G \, | \, A \, | \, \O_G} \; ,
\end{equation}
\begin{equation}\label{eq:uncertainty}
  \sum_{j=1}^n w_j \ = \ \ip{\O_G \, | \, A^2 \, | \, \O_G} -
  \ip{\O_G \, | \, A \, | \, \O_G}^2 \; ,
\end{equation}
and
\begin{equation}
  w_j \ = \ \lim_{E \rightarrow \eps_j} (\eps_j - E) \gamma(E)
  \ = \ {1 \over \ip{\O_G|(A-\eps_j)^{-2}|\O_G}} \; .
\end{equation}
\end{subequations}
The first two equalities may be verified by expanding $\gamma$ in
a Laurent series around $0$.
\end{enumerate}

\section{Examples and Applications}\label{sec:examples}
\subsection{Splitting the spectrum of the Laplacian on
$\Z^d$}\label{sec:example1} 
\begin{figure}[tbp]
\leavevmode
\centering \epsfysize=1.9in \epsffile{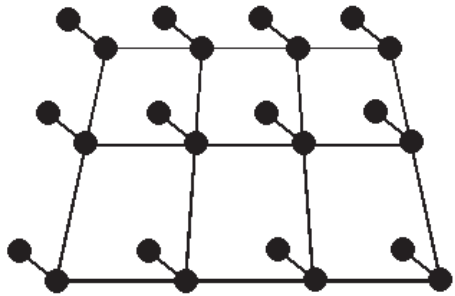}
\caption{\label{fig:example1} 
A portion of the graph discussed in \S\ref{sec:example1} 
 in the case $d=2$.
} 
\end{figure}

For a simple example of the
phenomenon, consider the operator 
$H = -\Delta_{\Z^d \triangleleft G}$ 
where $G$ is the graph consisting of two vertices
$V(G) = \{\O_G, 1_G\}$ and a single edge $E(G) = \{\{\O_G,1_G\}\}$
(see figure \ref{fig:example1}). 
As described at the end of \S \ref{sec:decoration},
$H$ is of the form
\eqref{eq:defnofH} with $H_o = -\Delta_{\Z^d}$ and $A=-\Delta_G$.

The spectrum of 
$H_{o}=-\Delta_{\Z^d}$
is
\begin{equation}
  \sigma(H_{o}) = [0,4 d]
\end{equation}
and all the associated spectral measures are purely absolutely
continuous.  
In this case, the function $\gamma$ is easy to calculate:
\begin{equation}
  \gamma(E) \ = \ - \left [ {1-E \over (1-E)^2 -1 }\right ]^{-1}
    \ = \ E \ - \ 1 \ + \ {1 \over 1 - E} \; ,
\end{equation}
and the vector $\left |\O_G \right >$ is cyclic. Thus
\begin{equation}
\begin{split}
  \sigma(H) \ =& \
  \left \{ E  \mid 0 \ \le \ E \ - \ 1 \ + \ {1 \over 1 - E} \ \le \ 4 d
   \right \} \\ =& \ \left [ 0, 1 + 2d - \sqrt{1 + 4d^2 } \right ] \cup
   \left [ 2, 1 + 2d + \sqrt{1 +  4d^2} \right ] \; ,
  \end{split}
\end{equation}
and the spectral measures are purely absolutely continuous.

\subsection{An example in which $\left | \O_G \right >$ is not cyclic.}
\label{sec:example2} 
\begin{figure}[bpt]
\leavevmode
\centering \epsfysize=1.9in \epsffile{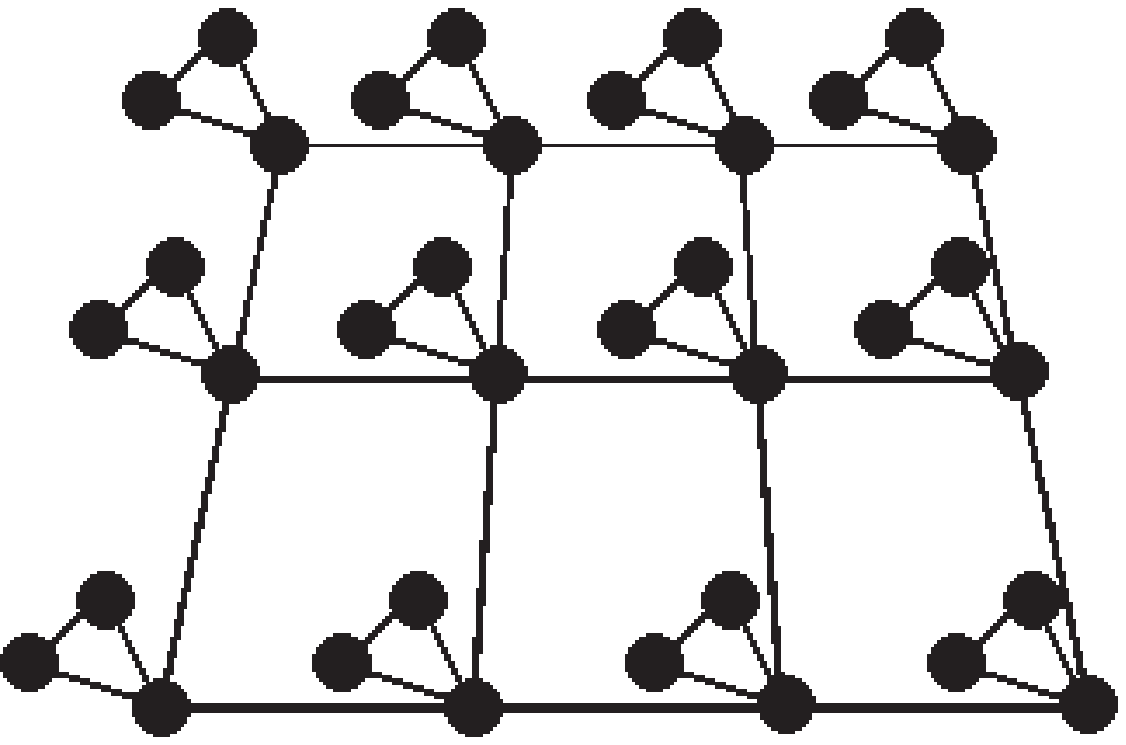}
\caption{\label{fig:example2}A situation in which $\left |\O_G
\right >$ is not cyclic (\S \ref{sec:example2}).}
\end{figure}

Consider now the discrete Laplacian on the
graph $\Z^d \triangleleft G$ where $G$ is the fully connected
graph with three vertices $V(G) = \{\O_G,1_G,2_G\}$ (see figure
\ref{fig:example2}).

The involution $R$ on $\ell^2(G)$ obtained by interchanging $\left
| 1_G \right >$ and $\left | 2_G \right >$ commutes with
$\Delta_G$. Hence $\Delta_G$ leaves invariant the subspaces
$V_{+(-)}$ of functions which are symmetric (anti-symmetric) with
respect to this involution.  A non-normalized basis of
simultaneous eigenfunctions for $R$ 
 and $(-\Delta_G ) $  consists of: $\left
|\O_G \right
> + \left |1_G \right > + \left |2_G \right > \in V_+$ with
eigenvalue $0$ , $2 \left |\O_G \right > - \left |1_G \right > -
\left |2_G \right > \in V_+ $ with eigenvalue $3$ , and $ \left
|1_G \right > - \left |2_G \right > \in V_- $ with eigenvalue $3$.

Using this basis, it is easy to see that $\left |\O_G \right >$ is
a cyclic vector for the restriction of $\Delta_G$ to $V_+$.
Furthermore, we can calculate $\gamma$:
\begin{equation}
  \gamma(E) \ = \ {- 1 \over {2 \over 3} {1 \over 3 - E} + {1
  \over 3}{1 \over 0 - E}} \ = \ E - 2 + {2 \over 1 - E} \; ,
\end{equation}
and note that $\sigma(-\Delta_G|_{V_-}) = \{3\}$. Thus,
\begin{equation}
  \begin{split}
    \sigma(-\Delta_{\Gamma \triangleleft G}) \ =& \ \{3\} \cup \left \{ E \mid
    0 \le {E - 2 + {2 \over 1 - E} } \le 4 d \right \} \\
    =& \left [ 0, \eps^- \right ] \cup \left [ 3, \eps^+ \right ] \; ,
  \end{split}
\end{equation}
where $\eps^{\pm}$ are the solutions to
\begin{equation}
  E - 2 + {2 \over 1 - E} \ = \ 4d \; ,
\end{equation}
with $\eps^- < 1$ and $\eps^+ > 1$.  The spectrum is purely
absolutely continuous except for the presence of an infinitely
degenerate eigenvalue at $E=3$.

\subsection{Persistence of band edge localization}
The operator $H$ may include disorder, in the form of
a random potential at the sites of $\Gamma_{o}$.
It is generally expected that in such a situation the spectrum of
 $H_{o}$ will exhibit Anderson localization
 ({\it i.e.}, dense pure point spectrum) at all the spectral edges.
(This has
been rigorously shown to be true in various situations~
\cite{FiKl,BCH,KSS,Klp,ASFH}).
Let us note that the mechanism
of gap creation via graph decorations preserves such band edge
localization, even if the randomness is not introduced at all the
sites of the decorated graph.

\newpage 


\end{document}